\documentclass[conference]{IEEEtran}
\usepackage{amsmath,amssymb,amsfonts}
\usepackage{graphicx}
\usepackage{subcaption}
\usepackage{bm}
\usepackage{booktabs}
\usepackage{cite}
\usepackage[hidelinks,bookmarks=false,pdfpagemode=UseNone]{hyperref}

\newcommand{\CN}{\mathcal{CN}}
\newcommand{\Th}{\bm{\Theta}}
\newcommand{\B}{\mathbf{B}}
\newcommand{\W}{\mathbf{W}}
\newcommand{\Pm}{\mathbf{P}}
\newcommand{\x}{\mathbf{x}}
\newcommand{\y}{\mathbf{y}}
\newcommand{\z}{\mathbf{z}}
\newcommand{\avec}{\mathbf{a}}
\newcommand{\I}{\mathbf{I}}

\usepackage{xcolor}

\begin{document}

\title{Goal-Oriented Wave Computing for Direction-of-Arrival Estimation}

\author{\IEEEauthorblockN{Mert Kalfa}
\IEEEauthorblockA{Faculty of Engineering and Natural Sciences, Sabanc{\i} University, Istanbul, T\"urkiye\\
Email: mert.kalfa@sabanciuniv.edu}
}

\maketitle

\begin{abstract}
Programmable metasurfaces are arrays of elements whose electromagnetic response can be tuned electronically. They can process signals as a wave propagates through or reflects off them, an approach known as wave-domain analog computing, or wave computing. The design strategy is typically indirect: the surface is fitted to a prescribed linear operator, such as a discrete Fourier transform (DFT), and the result is read from output power measurements. Using direction-of-arrival (DoA) estimation as a canonical benchmark, we study what this indirection costs across three analog computing topologies: reflective reconfigurable intelligent surfaces (RIS), beyond-diagonal RIS (BD-RIS), and stacked intelligent metasurfaces (SIM). We first derive Cram\'er--Rao bounds for the complete physical estimation chain, including one that accounts for output detectors measuring only intensity. We then formulate a goal-oriented alternative where the surfaces are trained directly through the deployed readout, with no operator target. We observe that goal-oriented training lets sparsely connected or shallower surfaces reach the task performance of their more complex operator-fitted counterparts, despite nearly identical Fisher information. For fully connected BD-RIS and six-layer SIM, the advantage persists under one- and two-bit control. From $16$ to $64$ elements, the connectivity a beyond-diagonal RIS needs grows with the aperture, and the depth saving of a stacked intelligent metasurface shrinks. Under tuner and per-layer loss, the goal-oriented designs mostly keep their advantage over DFT fits of the same hardware.
\end{abstract}

\begin{IEEEkeywords}
reconfigurable intelligent surfaces, beyond-diagonal RIS, stacked intelligent metasurfaces, wave-domain analog computing, direction-of-arrival estimation, Cram\'er--Rao bound
\end{IEEEkeywords}

\section{Introduction}
A \emph{programmable metasurface} is an arrangement of sub-wavelength elements whose electromagnetic response (e.g., the phase imparted on an impinging wave) can be tuned electronically. As a wave reflects off or passes through it, the field at any point is a weighted superposition of the re-radiated element fields. The surface thus applies a configurable linear transformation to the incident field at the speed of light, with almost no computing energy beyond that of holding the configuration. Performing signal processing \emph{in the wave domain} this way is termed wave-domain, in-wave, or analog electromagnetic computing \cite{Silva2014,Estakhri2019,An2023SIM,InWave2026}. We refer to it simply as \emph{wave computing} throughout this paper.

Three hardware families dominate the current literature. A \emph{reconfigurable intelligent surface} (RIS) is typically a single reflective layer whose elements are individually tunable, so its action can be described by a diagonal matrix of reflection coefficients \cite{DiRenzo2020}. A \emph{beyond-diagonal RIS} (BD-RIS) additionally provides tunable circuit interconnections \emph{between} elements. This makes the surface response a full matrix and enlarges the set of transformations one surface can realize \cite{BDRISreview2025,NeriniClerckx2024}. A \emph{stacked intelligent metasurface} (SIM) cascades several transmissive programmable layers separated by short free-space propagation, emulating the layers of a neural network \cite{An2023SIM}.

Across the families described above, the prevailing design approach is indirect and proceeds in two steps. First, the surface configuration is optimized so that the end-to-end response of the system approximates a prescribed target operator, such as a discrete Fourier transform (DFT), a beamforming matrix, or a general matrix-vector product \cite{An2023SIM,Silva2014}. Second, the result of the computation is read out from the output aperture. In reported wave-domain direction-of-arrival (DoA) systems, the surface is fitted to a DFT and the angle estimate is the index of the strongest output power bin \cite{An2024DoA}. The output detectors typically measure only \emph{intensity} (received power), referred to as \emph{intensity readout} throughout this paper. Task performance is then inherited from the operator fit rather than optimized directly.

Task-aware designs do exist, but they digitize on the \emph{data side}. Semantics-aware SIM uses the surface as a communication encoder feeding a digital receiver \cite{SemComSIM2024,MultimodalSemCom2025}. Dynamic metasurface antennas and diffractive meta-neural networks optimize the surface as an information-maximizing measurement front end whose outputs are processed by substantial algorithms \cite{TapieDelHougne2026,ADCe2e2026,DMNN2025}. In this work, we place the conversion on the \emph{answer side}: with the argmax readout, the analog chain of surface, propagation, and power-only detection performs the entire inference in a single snapshot. We train the surface \emph{as the estimator} and benchmark it against its own estimation-theoretic limits. DFT-fitted wave-domain DoA receivers similar to \cite{An2024DoA} serve as our baseline. Three recent works are closest. One names \emph{task-oriented wave processing} and lists the task-to-configuration mapping as an open challenge~\cite{TaskWave2026}. One trains SIM configurations with a digital layer for DoA and bounds them with an amplitude-only CRB~\cite{EMNN2026}. One finds that, with the input encoded in the surface configuration, strong coupling lets a single layer perform well~\cite{Expressivity2026}. Here, the input is the incident wavefront, the inference ends at an answer-side readout, and all fabrics share the same bounds.

This paper focuses on the design and training of a surface whose analog chain performs the entire inference for a canonical estimation task: single-source DoA estimation. The aperture itself acts as the estimator: a plane wave enters, and the index of the strongest output detector is the angle estimate. Our starting observation is that the surface, the propagation around it, and the intensity readout together form a complete \emph{physical estimator}, and should be analyzed and trained as one. Our contributions are summarized as follows:
\begin{itemize}
\item \textbf{Framework:} A goal-oriented formulation of wave computing over multiport network models of RIS, BD-RIS, and SIM. The models include reciprocity, wiring-limited BD-RIS connectivity (number of connected neighbors), and mutual coupling. The training objective is defined directly on the deployed estimator, with analytic gradients through the intensity readout (Secs.~\ref{sec:model},~\ref{sec:training}).
\item \textbf{Analytical Bounds:} Cram\'er--Rao bounds (CRBs) for the wave-domain DoA task, including an intensity readout bound that accounts for phase-insensitive detection. A resulting ``readout ladder'' separates information limits from estimator limits (Sec.~\ref{sec:bounds}).
\item \textbf{Observations:} Numerical evidence that (i) the accuracy of the standard fit-then-read pipeline (as used in \cite{An2024DoA} and in our baselines) is limited by the readout grid (output angular spacing) rather than by the information available in the wave, as the bounds quantify; (ii) goal-oriented training shifts the accuracy-per-hardware trade-off, allowing sparsely connected or shallower surfaces to reach the performance of more complex operator-fitted ones, with a depth saving that holds up to $N=32$ elements and shrinks at $N=64$; and (iii) surfaces with nearly identical Fisher information can differ by up to $3.6\times$ in task accuracy (Sec.~\ref{sec:results}).
\end{itemize}
The scenario is restricted on purpose. A linear array, one source, and one snapshot define a benchmark with analytical bounds and a known readout floor. A difference in accuracy can then be traced to the hardware or its training rather than to the scene. The paper thus focuses on the three fabrics, the complexity of each fabric, the training objective, and the effect of control quantization. The formulation carries over to planar apertures and multi-snapshot readouts, while the numerical values apply to the cases studied here. The fabrics under study are lossless and the same model is used in training and test, except in Sec.~\ref{sec:results-loss}, where we analyze the effects of lossy hardware on lossless designs. The hardware comparisons count tunable elements.

Throughout the paper, lowercase and uppercase boldface denote vectors ($\avec$) and matrices ($\mathbf{A}$); $(\cdot)^{T}$, $(\cdot)^{H}$, and $(\cdot)^{*}$ denote transpose, conjugate transpose, and complex conjugate; $\|\cdot\|$ is the Euclidean norm for vectors and the Frobenius norm for matrices; $\I$ is the identity matrix; $\CN(\bm{\mu},\mathbf{C})$ denotes the circularly symmetric complex Gaussian distribution; $\circ$ is the Hadamard product; and $\lceil\cdot\rceil$ is the ceiling function. Time-harmonic fields follow the $e^{j\omega t}$ convention (suppressed throughout), under which a wave propagating along the unit vector $\hat{\mathbf{k}}$ carries the spatial phase $e^{-jk\,\hat{\mathbf{k}}\cdot\mathbf{r}}$.

\section{System and Surface Models}\label{sec:model}
This section develops the physical model of the paper. It gives the geometry and end-to-end operator shared by all three fabrics (Sec.~\ref{sec:geom}), their multiport network descriptions (Sec.~\ref{sec:fabrics}), and a reachability constraint that reciprocity imposes on the reflective fabrics (Sec.~\ref{sec:reach}).
\begin{figure*}[!t]
\centering
\includegraphics[width=0.9\textwidth]{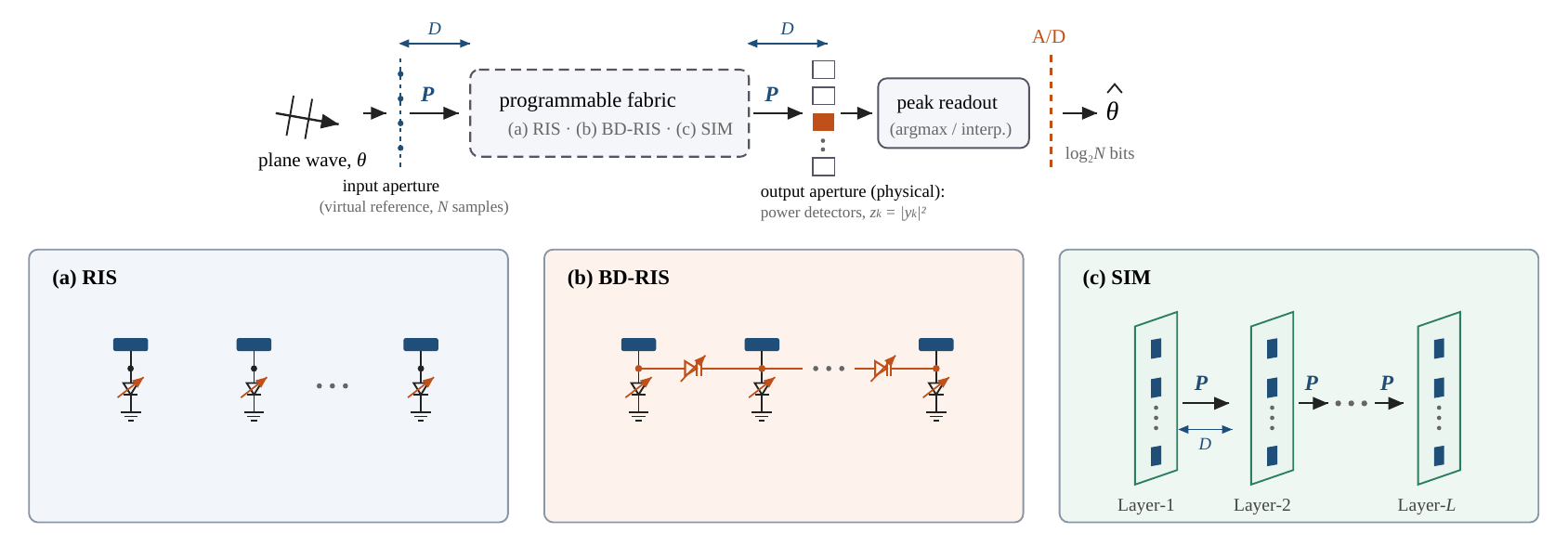}
\caption{Wave-domain DoA estimation. Top: the estimation chain. A plane wave sampled on the input aperture propagates to the programmable fabric and on to the output aperture of power detectors. The peak readout yields $\hat\theta$, so only $\lceil\log_2 N\rceil$ bits cross the A/D boundary. Bottom, (a) RIS, independent per-element tuners (varactor or PIN diode); (b) BD-RIS with tunable inter-element susceptances ($k$ nearest neighbors, $k{=}1$ shown); (c) SIM, $L$ transmissive layers of per-element tuners. Reflective (a,b) are drawn unfolded; the transmissive SIM has physically distinct apertures.}
\label{fig:system}
\end{figure*}

\subsection{Geometry and end-to-end operator}\label{sec:geom}
We consider one unified model for all three surface topologies (Fig.~\ref{fig:system}). The system contains only two kinds of hardware: the programmable surface(s), and an $N$-element power-detector \emph{output aperture} at standoff $D$. The incident field is referenced on a virtual, hardware-free \emph{input aperture} at distance $D$ on the illumination side, sampled at $N$ points with half-wavelength spacing. Elements lie on the $x$-axis and the planes stack along $+z$, with incidence angle $\theta$ measured from the array normal. A unit-amplitude plane wave from $\theta$ produces $\x = \avec(u)/\sqrt{N}$, $u = \sin\theta$, $a_n(u) = e^{j\pi n u}$, $n = 0,\dots,N-1$. The wave propagates to the surface, is transformed, and propagates to the output aperture, where each detector reports only its power,
\begin{equation}
z_k \;=\; |y_k|^2, \qquad \y = \W\x + \mathbf{n}, \label{eq:readout}
\end{equation}
with $\mathbf{n}\sim\CN(\mathbf{0},\sigma^2\I)$ modeling receiver noise and $\W$ the end-to-end operator realized by the computing fabric. The noise enters at the detection stage, as the passive analog chain contains no active components ahead of the detectors. The baseline fabrics and coupling networks are lossless (Sec.~\ref{sec:fabrics}); hence, $\W$ is unitary. Sec.~\ref{sec:results-loss} challenges this assumption. The linear array lattice and half-wavelength spacing are adopted for concreteness. The model extends directly to planar apertures, while denser spacings need a propagation model that keeps the mode amplitudes.

Each free-space hop is the same matrix $\Pm$, built from the free-space Green's function between planes at separation $D$ (the aperture--surface standoff for the reflective fabrics, drawn unfolded; or the layer spacing for SIM), with $x_m=(m-\tfrac{N-1}{2})\tfrac{\lambda}{2}$:
\begin{equation}
G_{mn} = \frac{e^{-jkR_{mn}}}{R_{mn}}, \qquad R_{mn} = \sqrt{D^2 + (x_m - x_n)^2},
\label{eq:kernel}
\end{equation}
$k = 2\pi/\lambda$. Treating propagation and surfaces as \emph{lossless} and \emph{reciprocal}, we keep the phase-only factor of this complex-symmetric kernel via a Takagi factorization $\mathbf{G} = \mathbf{U}_t\bm{\Sigma}\mathbf{U}_t^{T}$. This gives $\Pm = \mathbf{U}_t\mathbf{U}_t^{T}$, which is symmetric and unitary by construction. This idealization gives every mode of the kernel unit gain. Thus it neglects the spillover of a finite aperture: for $N=16$ and $D=15\lambda$, only six of the sixteen singular values of $\mathbf{G}$ exceed $10\%$ of the largest. It also requires half-wavelength sampling, at which no sampled component is evanescent. Placing input and output apertures symmetrically then makes $\W=\Pm\bm\Gamma\Pm$ symmetric for a reciprocal surface ($\bm\Gamma$ its effective reflection, Sec.~\ref{sec:fabrics}), which Sec.~\ref{sec:reach} exploits.

The input/output apertures are physically distinct for the transmissive SIM but co-located for the reflective RIS/BD-RIS. There, we treat them as an idealization realizable bistatically (a receive array offset from the illumination) or via per-element circulators. A bistatic build adds an angle-dependent element-gain taper, which we omit here.

\subsection{Multiport surface models}\label{sec:fabrics}
Each programmable surface is modeled as an $N$-port reconfigurable network terminating an $N$-antenna array \cite{BDRISreview2025,NeriniClerckx2024}. For a lossless, reciprocal network parametrized by a real symmetric susceptance matrix $\B$ (entries $B_{mn}$ are the tunable reactive elements connecting ports $m$ and $n$), the surface scattering matrix is the Cayley transform
\begin{equation}
\Th = (Y_0\I - j\B)(Y_0\I + j\B)^{-1},
\label{eq:cayley}
\end{equation}
which is symmetric (reciprocity) and unitary (losslessness) for \emph{any} real symmetric $\B$, where $Y_0$ is the reference admittance.

For the reflective fabrics (RIS and BD-RIS), we model the surface as a $2N$-port network: $N$ \emph{radiation-side} ports (each element's interface to free space) and $N$ \emph{network-side} ports (the feed terminals closed by the tunable load network $\Th$). Partitioning its scattering matrix into $N\times N$ blocks,
\begin{equation}
\mathbf{S} = \begin{bmatrix}\mathbf{S}_{rr} & \mathbf{S}_{rn}\\ \mathbf{S}_{nr} & \mathbf{S}_{nn}\end{bmatrix},
\label{eq:blocks}
\end{equation}
where subscript $r$ denotes the radiation side and $n$ the network side. Here, $\mathbf{S}_{rr}$ is the structural (radiation-side) reflection, $\mathbf{S}_{rn} = \mathbf{S}_{nr}^{T}$ is the link between the two sides (transmission or reception), and $\mathbf{S}_{nn}$ is the network-side reflection block, whose off-diagonal entries encode the \emph{mutual coupling} between elements. Terminating the network side with $\Th$ gives the effective radiation-side reflection
\begin{equation}
\bm{\Gamma} = \mathbf{S}_{rr} + \mathbf{S}_{rn}\,\Th\,(\I - \mathbf{S}_{nn}\Th)^{-1}\mathbf{S}_{nr},
\label{eq:gamma}
\end{equation}
which reduces to $\bm{\Gamma} = \Th$ in the idealized decoupled case ($\mathbf{S}_{rr} = \mathbf{0}$, $\mathbf{S}_{rn} = \mathbf{S}_{nr} = \I$, $\mathbf{S}_{nn} = \mathbf{0}$). For the coupling-aware case, we compute $\mathbf{S}_{nn}$ from a classical thin-dipole mutual-impedance model~\cite{Carter1932},
\begin{equation}
\mathbf{S}_{nn} = (\bar{\mathbf{Z}} - \I)(\bar{\mathbf{Z}} + \I)^{-1}, \qquad \bar{\mathbf{Z}} = \mathbf{Z}/Z_0,
\label{eq:mc}
\end{equation}
with $Z_{mn}$ the mutual impedance between parallel thin half-wave dipoles at spacing $|x_m - x_n|$, pre-computed in closed form via~\cite{Carter1932}. With $Z_{mm}\approx 73.1 + j42.5\,\Omega$ and $Z_0 = 73\,\Omega$, we obtain $\|\mathbf{S}_{nn}\|_2 = 0.76$ for $N=16$.

Mutual-coupling-aware BD-RIS models derive all blocks from one impedance description~\cite{BDRISMC2023,GradoniDiRenzo2021}. Here we use the lossless reciprocal completion
\begin{equation}
\mathbf{S}_{rr} = -\mathbf{S}_{nn}^{*},\qquad \mathbf{S}_{rn} = \mathbf{S}_{nr}^{T} = \big(\I - \mathbf{S}_{nn}^{H}\mathbf{S}_{nn}\big)^{1/2},
\label{eq:completion}
\end{equation}
with the Hermitian square root. For a symmetric $\mathbf{S}_{nn}$ with $\|\mathbf{S}_{nn}\|_2\le 1$, the matrix $\mathbf{S}$ in \eqref{eq:blocks} is then symmetric and unitary. Unitarity follows from the identity $\mathbf{S}_{nn}(\I-\mathbf{S}_{nn}^{H}\mathbf{S}_{nn})^{1/2} = (\I-\mathbf{S}_{nn}\mathbf{S}_{nn}^{H})^{1/2}\mathbf{S}_{nn}$. The completion reduces to the decoupled case for $\mathbf{S}_{nn}=\mathbf{0}$. A lossless reciprocal $2N$-port terminated by a lossless reciprocal $\Th$ gives a symmetric unitary $\bm{\Gamma}$ in \eqref{eq:gamma}, so the reflective $\W$ stays symmetric and unitary with coupling. The results in Sec.~\ref{sec:results} use the decoupled model.

The three fabrics are distinguished by the sparsity of $\B$ and by how the surface composes with $\Pm$ into $\W$. A diagonal $\B$ is a \emph{RIS} ($N$ tuners, $\W = \Pm\bm{\Gamma}\Pm$). A banded $\B$ of half-bandwidth $k$ is a wiring-limited \emph{BD-RIS} (each element wired to its $k$ nearest neighbors; full $\B$ is fully connected). It has the same reflective form $\W = \Pm\bm{\Gamma}\Pm$, with the reachable $\Th$ growing in $k$. A \emph{SIM} of $L$ transmissive diagonal layers $\bm{\Phi}_\ell$ realizes the cascade $\W = \Pm\,\bm{\Phi}_L\Pm\cdots\bm{\Phi}_1\Pm$.

\subsection{A reachability constraint from reciprocity}\label{sec:reach}
For the reflective fabrics, $\Th$ and $\Pm$ are symmetric, so $\W$ is too: \emph{a single passive reciprocal surface can only realize symmetric operators}. Let $\mathbf{T}\in\mathbb{C}^{N\times N}$ denote the normalized DFT matrix, $T_{mn}=e^{-j2\pi mn/N}/\sqrt{N}$ for $m,n=0,\dots,N-1$. Its output $m$ peaks at the bin angle $u_m=2m/N$, wrapped to $[-1,1)$. These bin angles form the grid of the readouts in Sec.~\ref{sec:bounds}. The DFT is symmetric unitary and is exactly reachable by a fully connected BD-RIS ($\Th^\star = \Pm^{-1}\mathbf{T}\Pm^{-1}$, with real symmetric $\B^\star$ from \eqref{eq:cayley}), whereas a generic (e.g.\ pseudo-inverse) beamformer is not. Target operators for the reflective fabrics (RIS and BD-RIS) must thus be symmetric. The DFT satisfies this condition, and we revisit it when choosing training objectives. A SIM with $L\ge 2$ is not bound by this constraint, because $\W^T$ reverses the layer order.

\section{Estimation-Theoretic Bounds}\label{sec:bounds}
We now ask how accurately \emph{any} estimator could recover $u$ from the physical observations. For the lossless baseline, the coherent bound depends only on the noise level $\sigma^2$ and the element count $N$, because a unitary $\W$ preserves the Fisher information of a coherent observation. We investigate loss effects in Sec.~\ref{sec:results-loss}. The intensity bound also depends on $\W$, through the output field. Each bound is summarized by its median over $u\sim\mathcal{U}(-0.9,0.9)$. Table~\ref{tab:ladder} lists the values for the DFT operator at the operating point, and Sec.~\ref{sec:results} evaluates the intensity bound of the trained surfaces.

\subsection{Coherent bound}
If the receiver could observe the complex output $\y$ in~\eqref{eq:readout} coherently, then the standard single-source array result \cite{StoicaNehorai1989} gives
\begin{equation}
\mathrm{CRB}_u \;=\; \frac{6\,\sigma^2}{|g|^2\,\pi^2 N(N^2-1)},\qquad
\mathrm{CRB}_\theta = \frac{\mathrm{CRB}_u}{1-u^2},
\label{eq:crbcoh}
\end{equation}
where $|g|^2 = 1/N$ is the per-element signal power. We summarize each bound by the median over $u$ of the per-angle root-CRB $\sqrt{\mathrm{CRB}_\theta(u)}$ (the RMS is undefined, as the intensity bound below diverges at bin centers).

\subsection{Intensity-only readout bound}
Under the \emph{intensity readout}, the physical estimator observes only the powers $z_k$ in \eqref{eq:readout}. The outputs $y_k\sim\CN(\mu_k,\sigma^2)$ are independent for any $\W$, with $\bm{\mu} = \W\avec(u) g$. Each $z_k$ therefore follows a scaled noncentral-$\chi^2$ law with signal power $\lambda_k = |\mu_k|^2$ (noncentrality parameter $2\lambda_k/\sigma^2$). The exact per-detector Fisher information follows from the noncentral-$\chi^2$ score; its high-SNR (Gaussian) approximation is
\begin{equation}
J(u) \;\approx\; \sum_{k=0}^{N-1}\frac{\big(\lambda_k'(u)\big)^2}{\sigma^2\big(\sigma^2 + 2\lambda_k(u)\big)},
\qquad \lambda_k' = 2\,\mathrm{Re}\{\bar\mu_k \mu_k'\}.
\label{eq:fimint}
\end{equation}
Two properties follow. First, for the DFT operator, away from the output-bin-center angles the bound is close to the coherent bound (median ratio $1.008$). At high SNRs, discarding phase thus costs little. Second, \emph{at} an angle aligned exactly with an output bin, every $\lambda_k'$ vanishes and the bound diverges. This phase-retrieval-like degeneracy is exactly measure-zero only in the noiseless limit. At $20$\,dB per-element SNR it broadens into $\approx\!15$ narrow neighborhoods (for $N=16$) around the output-bin angles, where the intensity/coherent root-CRB ratio exceeds $2$. They occupy $\approx\!2.5\%$ of the angular range, which is why the summary is a median over $u$. The Gaussian form \eqref{eq:fimint} matches the exact noncentral-$\chi^2$ Fisher information away from these neighborhoods with a median error of $0.7\%$ and a maximum error of $5.6\%$. A maximum-likelihood Monte-Carlo estimate at $u=0.3$ and $20$\,dB is within $0.4\%$ of the intensity bound.

\subsection{The readout ladder}
We now make the deployed readouts explicit. Let $k_0$ denote the peak bin (the index of the largest output-bin power $z_k$). The \emph{argmax} readout returns the peak-bin angle $\hat u_{\rm arg}=u_{k_0}$, while the \emph{3-bin parabolic} readout fits a quadratic to the peak and its two neighbors,
\begin{equation}
\hat u_{\rm par}=u_{k_0}+\tfrac{\Delta u}{2}\,\frac{s_{k_0-1}-s_{k_0+1}}{s_{k_0-1}-2s_{k_0}+s_{k_0+1}},\quad \Delta u=\tfrac{2}{N},
\label{eq:readouts}
\end{equation}
a standard refinement of the peak location at almost no cost. The samples are either the powers $s_k=z_k$ or their logs $s_k=\ln z_k$. The log fit, applied digitally at identical A/D cost, lowers the grid-bias floor further (Table~\ref{tab:ladder}). These readouts are used throughout.

The bounds expose a hierarchy of estimators that share one physical front end, the ideal DFT operator. This ladder is a readout property and holds for any $\W$ realizing the DFT under either training scheme. Table~\ref{tab:ladder} lists it with what each level must send across the analog-to-digital boundary.
\begin{table}[t]
\vspace*{4pt}
\caption{The readout ladder as the analog--digital (A/D) partition ($N{=}16$, $20$\,dB per element SNR). Argmax/parabolic rows are noise-free grid RMS; ML on $\z$ is a non-deployable reference that needs the full power vector and a digital search. It is obtained by Monte Carlo at $u=0.3$, where the intensity bound is $0.073^\circ$. The coherent row is a fully digital receiver ($N$ RF chains, single snapshot, ML); its Monte-Carlo RMSE meets the RMS coherent bound over $u$ ($0.089^\circ$). The bounds are medians for the DFT operator.}
\label{tab:ladder}
\centering
\footnotesize
\setlength{\tabcolsep}{5.06pt}
\begin{tabular}{@{}llc@{}}
\toprule
readout & crosses A/D & RMSE\\
\midrule
argmax ($N$ bins) & answer ($\lceil\log_2 N\rceil$ bits) & $2.56^\circ$\\
3-bin parabolic (power) & 3 power samples & $1.78^\circ$\\
3-bin parabolic (log) & 3 power samples & $1.10^\circ$\\
ML on $\z$ & $N$ powers $+$ digital est. & $0.074^\circ$ at $u=0.3$\\
coherent ML (digital) & $N$ RF chains & $0.089^\circ$\\
\midrule
CRB (intensity / coherent) & & $0.083^\circ$ / $0.078^\circ$\\
\bottomrule
\end{tabular}
\end{table}
The argmax ``floor'' equals the RMSE of noise-free grid quantization ($\Delta u = 2/N$). The ladder is also a map of the A/D partition: each level moves the conversion boundary from the answer side toward the data side. The maximum-likelihood level is the regime of the task-aware front ends of \cite{TapieDelHougne2026,ADCe2e2026,DMNN2025}. The design question of Sec.~\ref{sec:training} is whether training can shape the response so that the cheapest (answer-side) readout extracts the available information.

\section{Goal-Oriented Training}\label{sec:training}
Section~\ref{sec:bounds} showed that the accuracy of the fit-then-read pipeline is set by the readout rather than by the information in the wave. This motivates configuring the surface to serve the deployed readout directly, instead of fitting an intermediate operator. This section defines such a goal-oriented objective on the deployed estimator (Sec.~\ref{sec:objective}) and derives its gradient through the physical fabric models (Sec.~\ref{sec:grad}).
\subsection{Objective defined on the deployed estimator}\label{sec:objective}
Let $\hat u(\z)$ denote either of the deployed estimators given in \eqref{eq:readouts}. The goal-oriented training objective drops the operator target entirely and penalizes task error directly, through the readout:
\begin{align}
\mathcal{L} \;=\;\; & \alpha\,\underbrace{\mathrm{CE}\big(\mathrm{softmax}(\beta\z),\, k^\star(u)\big)}_{\text{peak on the correct bin}} \nonumber\\[-2pt]
&+\; (1-\alpha)\,\underbrace{\big(\arcsin \hat u_{\rm par}(\z) - \arcsin u\big)^2}_{\text{$\theta$-domain error of the deployed readout}},
\label{eq:loss}
\end{align}
averaged over random angles \emph{and noise draws}. Here $\mathrm{softmax}(\beta\z)_k = e^{\beta z_k}/\sum_j e^{\beta z_j}$ is a differentiable soft peak with sharpness $\beta$. The cross-entropy $\mathrm{CE}(\mathbf p, k^\star)=-\log p_{k^\star}$ rewards placing the peak on the bin $k^\star(u)$ nearest the true angle. The term $\hat u_{\rm par}$ is the parabolic readout \eqref{eq:readouts} with the peak index held fixed during differentiation, so the loss is piecewise differentiable. The weight $\alpha\in[0,1]$ balances the two terms (we use $\alpha=0.5$): the cross-entropy keeps the peak on the correct bin, and the parabolic term sharpens the localization within it. The error is expressed in the angle domain, matching the reported metric in Sec.~\ref{sec:results}. This choice weights wide angles more heavily, where the CRB is also larger ($\mathrm{CRB}_\theta \propto 1/(1-u^2)$).

\subsection{Gradients through the physics}\label{sec:grad}
We collect a training batch of $M$ inputs into $\mathbf{X} = [\x_1,\dots,\x_M]$. The outputs are $\mathbf{Y} = \W\mathbf{X} + \mathbf{V}$, where $\mathbf{V}$ holds the noise samples. The columns of the element-wise intensities $\mathbf{Z} = |\mathbf{Y}|^2$ are the readout inputs $\z$ in \eqref{eq:loss}. The complex (Wirtinger) gradient of \eqref{eq:loss} with respect to the operator is
\begin{equation}
\partial\mathcal{L}/\partial\W^{*} = (\mathbf{C}\circ\mathbf{Y})\mathbf{X}^{H}, \qquad \mathbf{C} = \partial\mathcal{L}/\partial\mathbf{Z},
\label{eq:wgrad}
\end{equation}
where $\mathbf{C}$ collects the readout derivatives over the batch. This upstream gradient enters the same chain rules used for operator fitting (the Cayley differential for BD-RIS, layer backpropagation for SIM). Operator fitting uses Adam with $5000$--$9000$ iterations, and each SIM design keeps the best of ten random restarts. Goal-oriented training then runs Adam for $3000$ iterations ($4000$ with quantization, Sec.~\ref{sec:results-quant}), batch $64$, warm-started from the operator fit, with $\beta=20$ and $\alpha=0.5$. Sec.~\ref{sec:results-sweep} shows that neither the warm start nor the extra iterations explain the gain. The objective must be built on the \emph{deployed} readout. We do not use a soft-argmax surrogate $\sum_k u_k[\mathrm{softmax}(\beta\z)]_k$, because the optimizer can split energy across bins to satisfy the soft mean. The cross-entropy term keeps the peak on one bin.

\section{Numerical Results}\label{sec:results}

\begin{figure*}[t]
\centering
\begin{subfigure}[t]{0.3\textwidth}\centering\includegraphics[width=\linewidth]{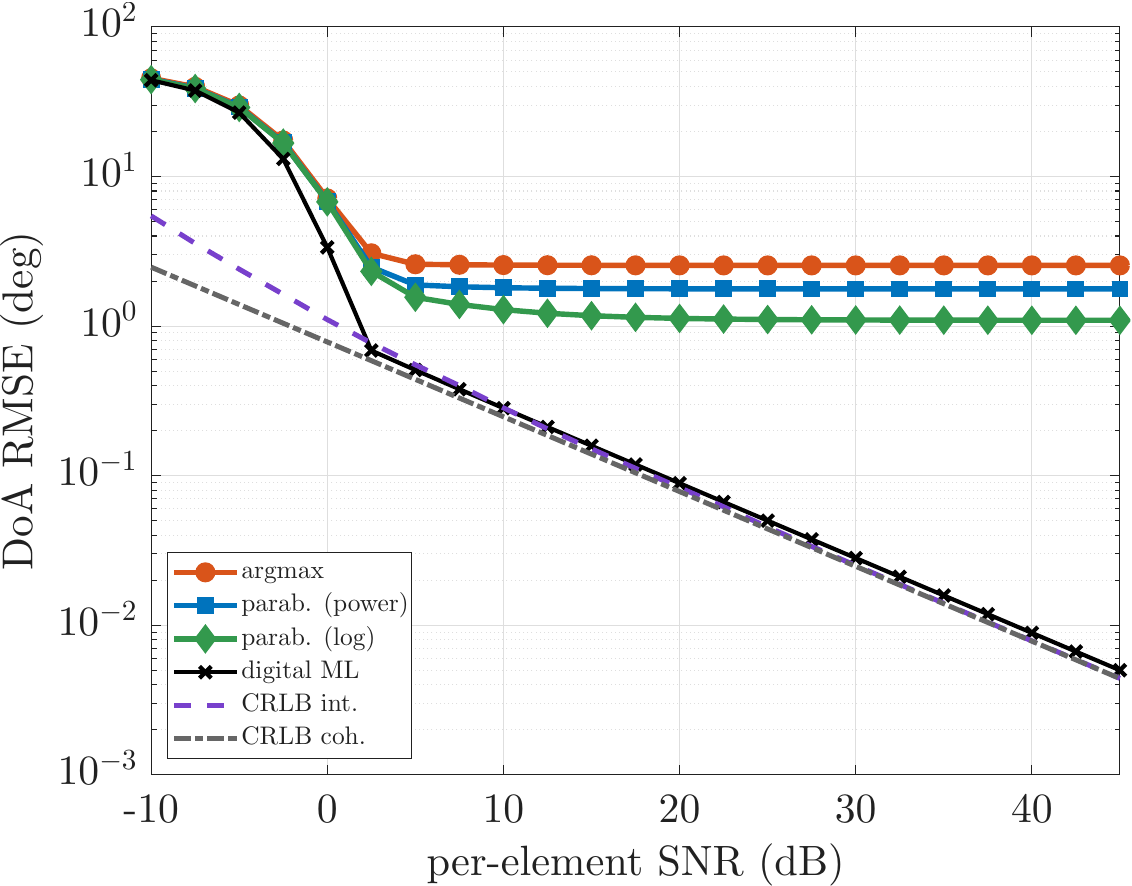}\caption{BD-RIS: SNR sweep}\label{fig:snr_bdris}\end{subfigure}\hfill
\begin{subfigure}[t]{0.3\textwidth}\centering\includegraphics[width=\linewidth]{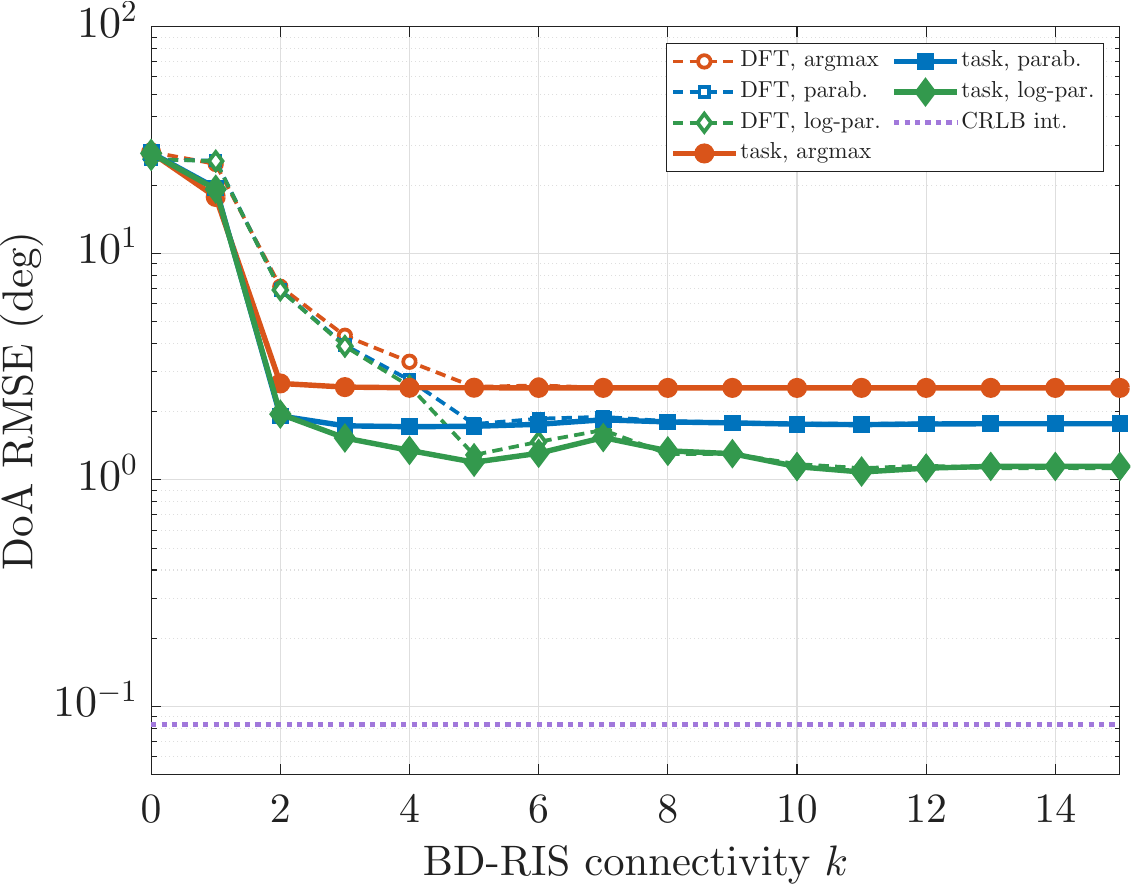}\caption{BD-RIS: connectivity $k$}\label{fig:sweep_bdris}\end{subfigure}\hfill
\begin{subfigure}[t]{0.3\textwidth}\centering\includegraphics[width=\linewidth]{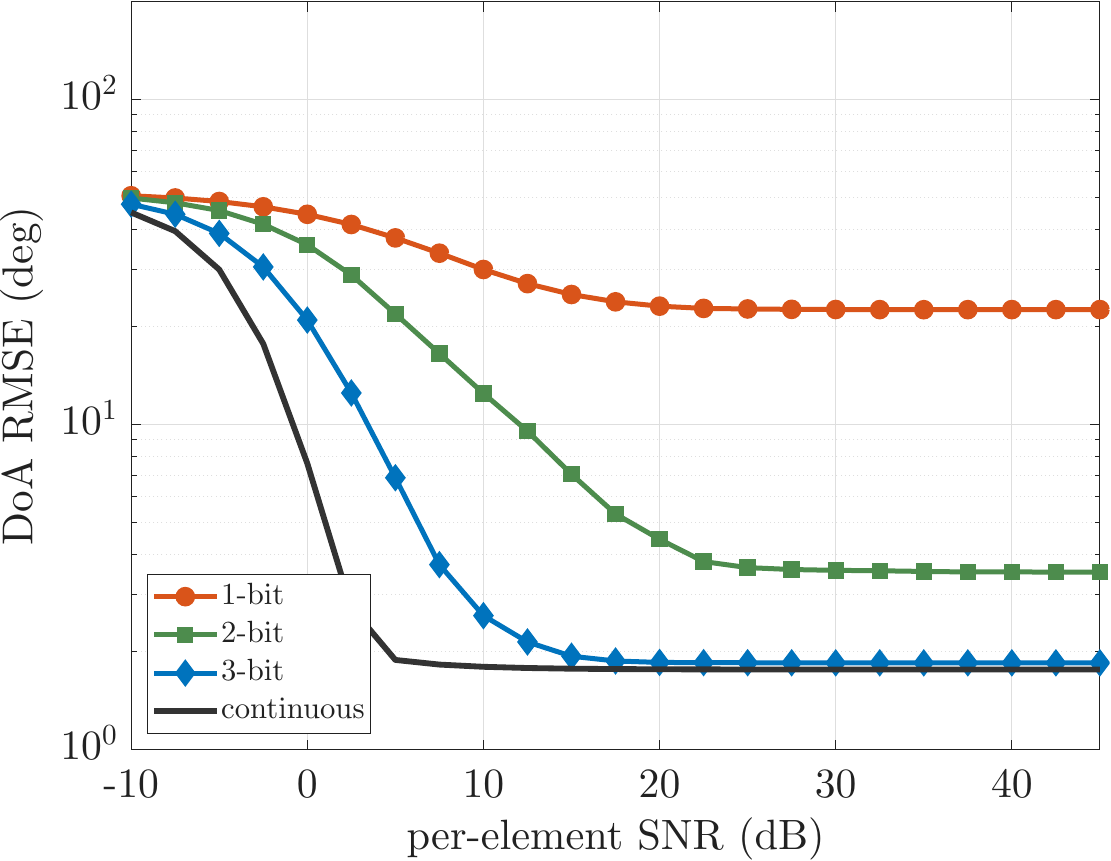}\caption{BD-RIS: $b$-bit control}\label{fig:qat_bdris}\end{subfigure}

\vspace{2pt}
\begin{subfigure}[t]{0.3\textwidth}\centering\includegraphics[width=\linewidth]{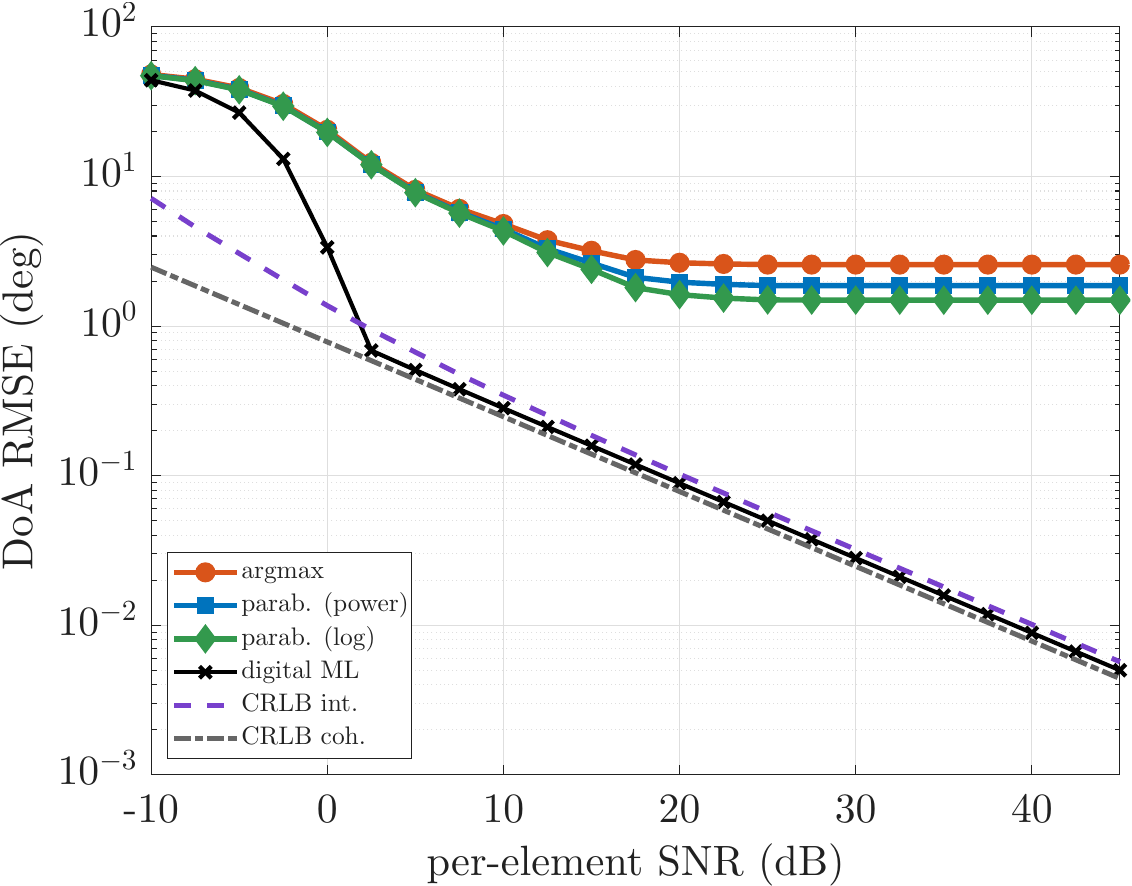}\caption{SIM: SNR sweep}\label{fig:snr_sim}\end{subfigure}\hfill
\begin{subfigure}[t]{0.3\textwidth}\centering\includegraphics[width=\linewidth]{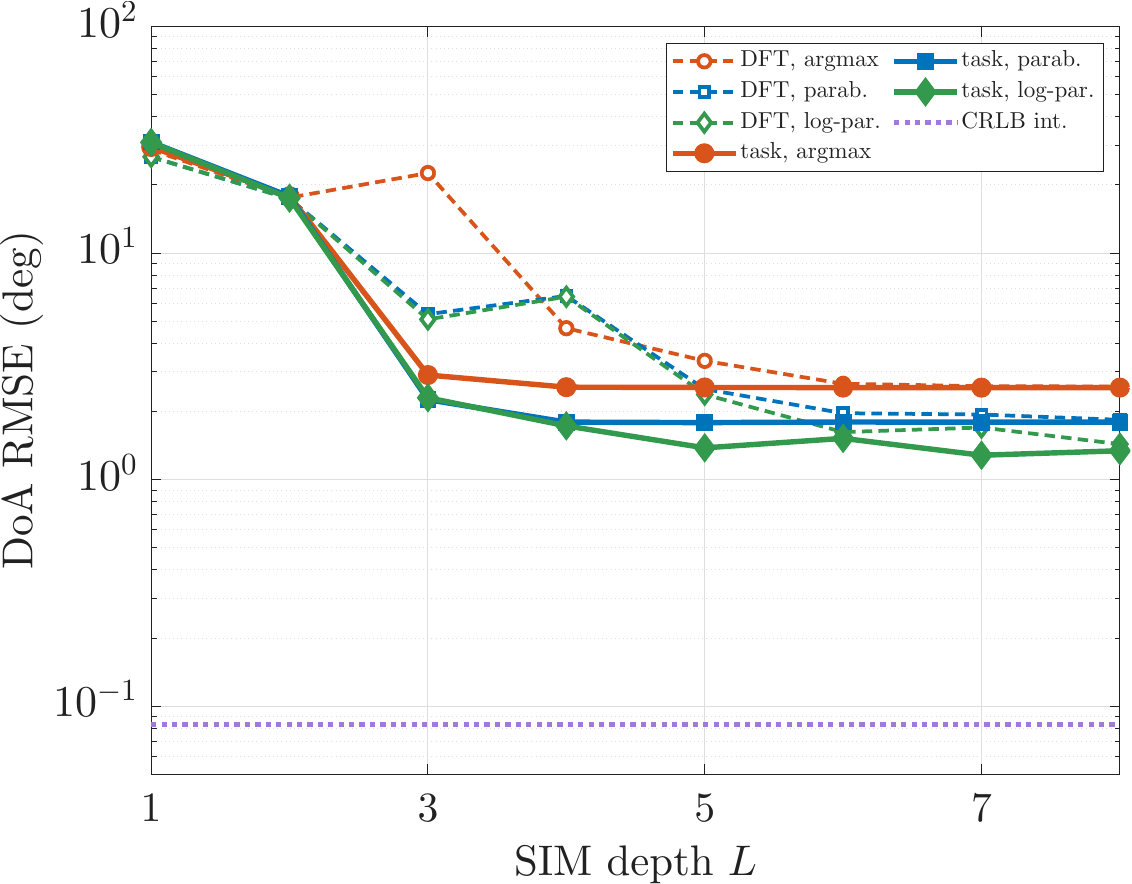}\caption{SIM: depth $L$}\label{fig:sweep_sim}\end{subfigure}\hfill
\begin{subfigure}[t]{0.3\textwidth}\centering\includegraphics[width=\linewidth]{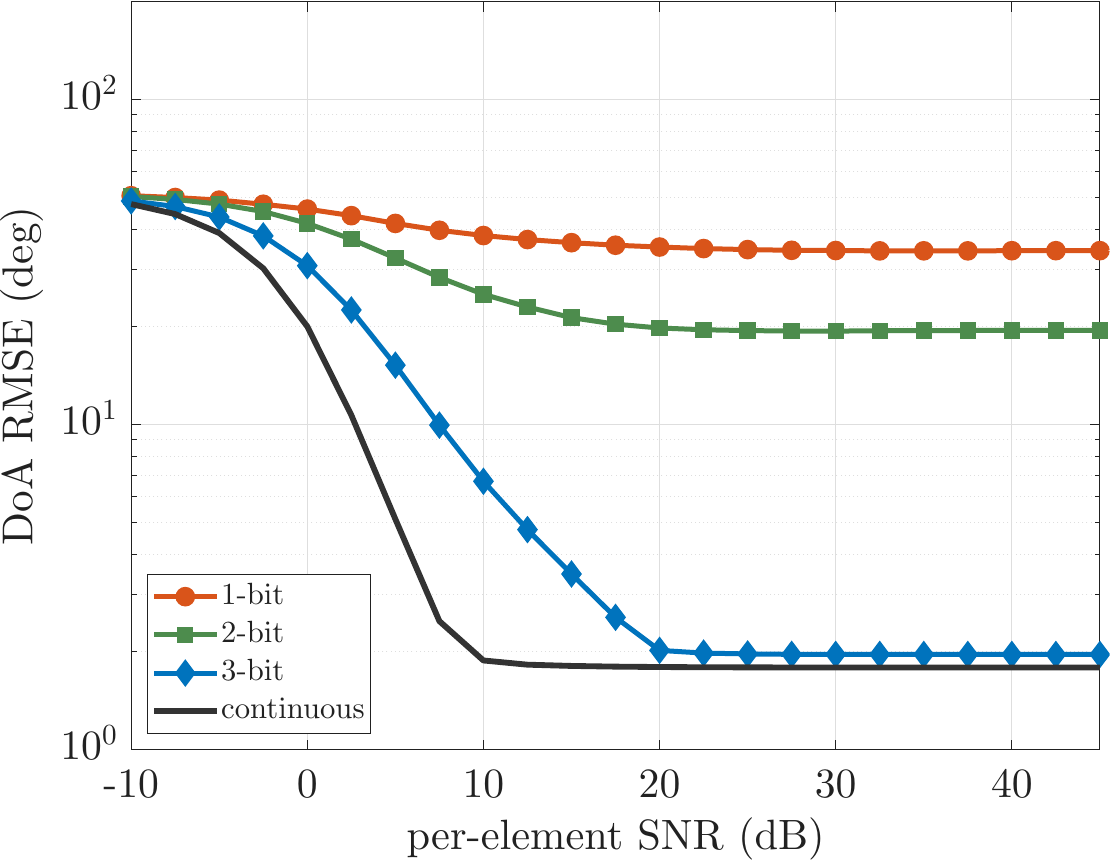}\caption{SIM: $b$-bit control}\label{fig:qat_sim}\end{subfigure}
\caption{Wave-computing DoA performance; top row BD-RIS, bottom row SIM. Left: RMSE vs.\ per-element SNR, for the argmax and power-/log-parabolic readouts with the corresponding CRBs and a fully digital ML receiver. Middle: DFT-fitted RMSE vs.\ goal-oriented RMSE training over the full connectivity/depth sweep, for three readouts with the intensity CRB. Right: goal-oriented $b$-bit control vs.\ SNR, continuous design given as a baseline.}
\label{fig:results}
\end{figure*}

In the following experiments, $N=16$ elements (Sec.~\ref{sec:results-n} varies $N$) and $60\,000$ Monte Carlo trials per point are used. The aperture--surface standoff is $D=15\lambda$. With an $8\lambda$-long aperture (a far-zone distance of $128\lambda$) the propagation is near-field, and since all lengths scale with $\lambda$ the results are frequency-independent. We report SNR \emph{per element}, $\mathrm{SNR}=\|\x\|^2/(N\sigma^2)$, which equals $1/(N\sigma^2)$ for the unit-norm input. The coherent array gain adds $10\log_{10}N\approx12$\,dB, so $20$\,dB per element is about $32$\,dB of array SNR. Except where a sweep is stated, the operating point is $20$\,dB per element. We chose it high to compare designs in the readout-limited regime. The reconfigurable elements are \emph{idealized}: continuous-valued and lossless (Sec.~\ref{sec:results-quant} relaxes this to $b$-bit control and Sec.~\ref{sec:results-loss} adds loss). Mutual coupling is off ($\mathbf{S}_{nn}=\mathbf{0}$). Each SIM layer is a transmissive surface with its own \emph{per-element} tuners, so a depth-$L$ SIM has $LN$ independent per-element phases.

\subsection{The readout floor and its SNR dependence}\label{sec:results-snr}
The left column of Fig.~\ref{fig:results} shows the SNR dependence of a strong design of each fabric (fully connected BD-RIS; $L=6$ SIM). At low SNR the readouts are noise-limited and track the bounds. Above $5$--$25$\,dB per element they \emph{saturate} at fixed, interpolator-dependent floors, while the CRB keeps falling. At the $20$\,dB operating point, the floors sit $14$ to $31\times$ above the intensity CRB, depending on the readout. Peak-picking and parabolic interpolation are biased by the output grid, so they do not reach the true angle even at infinite SNR. At high SNR, the readout is therefore the bottleneck rather than the information in the wave. A fully digital receiver with $N$ coherent RF chains and single-snapshot ML reaches the coherent bound, at the price of $N$ RF chains instead of $N$ power detectors. Averaging $T$ intensity snapshots multiplies the Fisher information by $T$ but does not lower the readout floors.

\subsection{Goal-oriented vs.\ operator-fitted training}\label{sec:results-sweep}
The middle column of Fig.~\ref{fig:results} sweeps the full BD-RIS connectivity $k=0,\dots,N-1$ (Fig.~\ref{fig:sweep_bdris}) and SIM depth $L=1,\dots,8$ (Fig.~\ref{fig:sweep_sim}) at $20$\,dB SNR. It compares DFT-fitted and goal-oriented training under three readouts (argmax, power- and log-parabolic). A BD-RIS with $k$-neighbor connectivity has $k(2N{-}k{-}1)/2$ tunable inter-element couplings in addition to the $N$ element tuners.

Under DFT fitting, the argmax RMSE of the BD-RIS reaches the grid floor at $k\!\approx\!5$ ($65$ of the $120$ full couplings), while a SIM needs $L\!\approx\!6$. At $N=16$, connectivity is thus a more hardware-efficient route than depth. Goal-oriented training brings sparsely connected and shallower fabrics to the performance of their more complex operator-fitted counterparts. With the argmax and power-parabolic readouts, the $k=2$ BD-RIS ($29$ couplings) comes within $10\%$ of the fully connected DFT-fitted result, and the $L=4$ SIM matches the $L=6$ stack. The DFT-fitted and goal-oriented $k=2$ surfaces carry nearly the same intensity-readout Fisher information (median $J(u)$ within $4\%$), yet their RMSE differs by $2.7\times$ (argmax) to $3.6\times$ (parabolic). The training gain therefore comes from aligning the intensity pattern with the deployed readout rather than from acquiring information. Training does not help every design. The diagonal RIS ($k=0$) is not improved. The $L=2$ SIM improves only with multiple starts ($17.6^\circ$ to $9.7^\circ$, best of three) and stays far above the floor.

Table~\ref{tab:budget} separates the objective from the budget and the warm start. A longer operator fit does not change its accuracy. Goal-oriented training from random initializations comes within $4\%$ of the warm-started designs and improves on them for the BD-RIS. The gain therefore comes from the task-oriented objective.
\begin{table}[t]
\caption{Training control at $N=16$ and $20$\,dB: RMSE for the argmax (arg.) and power-parabolic (par.) readouts. Cold start: BD-RIS $8000$ iterations; SIM ten starts of $9000$ iterations, best on a disjoint validation set.}
\label{tab:budget}
\centering
\small
\setlength{\tabcolsep}{11.68pt}
\begin{tabular}{@{}lcccc@{}}
\toprule
 & \multicolumn{2}{c}{BD-RIS, $k=2$} & \multicolumn{2}{c}{SIM, $L=4$}\\
\cmidrule(lr){2-3}\cmidrule(l){4-5}
training & arg. & par. & arg. & par.\\
\midrule
DFT fit & $7.12^\circ$ & $6.89^\circ$ & $4.66^\circ$ & $6.47^\circ$\\
DFT fit, $+3000$ it. & $7.09^\circ$ & $6.85^\circ$ & $4.66^\circ$ & $6.47^\circ$\\
goal-or., warm start & $2.66^\circ$ & $1.91^\circ$ & $2.56^\circ$ & $1.80^\circ$\\
goal-or., cold start & $2.61^\circ$ & $1.63^\circ$ & $2.62^\circ$ & $1.86^\circ$\\
\bottomrule
\end{tabular}
{}
\end{table}

\subsection{Array size}\label{sec:results-n}
Table~\ref{tab:nscale} repeats the sweeps for $N=16$, $32$, and $64$ at the same standoff and SNR. Each design is the best of three starts, selected on its own criterion (operator NMSE for DFT fitting, validation RMSE for goal-oriented training). For $N\ge 32$ at this standoff the kernel \eqref{eq:kernel} is numerically rank-deficient. We therefore compute its Takagi factor from the real symmetric embedding of $\mathbf{G}$ and complete it to a unitary $\Pm$.
\begin{table}[t]
\caption{Array size at $20$\,dB: smallest $k^\star$ and $L^\star$ with RMSE within $10\%$ of the fully connected DFT-fitted floor; a/b = argmax/parabolic.}
\label{tab:nscale}
\centering
\small
\setlength{\tabcolsep}{8.49pt}
\begin{tabular}{@{}cccccc@{}}
\toprule
 & floor & \multicolumn{2}{c}{BD-RIS $k^\star$} & \multicolumn{2}{c}{SIM $L^\star$}\\
\cmidrule(lr){3-4}\cmidrule(l){5-6}
$N$ & arg./par. & DFT fit & goal-or. & DFT fit & goal-or.\\
\midrule
16 & $2.55^\circ$/$1.77^\circ$ & 5 & 2 & 6/7 & 3/4\\
32 & $1.30^\circ$/$0.91^\circ$ & 10/11 & 7/9 & 9/10 & 5/6\\
64 & $0.67^\circ$/$0.46^\circ$ & 17/19 & 11/14 & 9/12 & 9\\
\bottomrule
\end{tabular}
{}
\end{table}

Table~\ref{tab:nscale} shows that goal-oriented training never needs more hardware than DFT fitting, but the saving depends on $N$. For the BD-RIS, $k^\star$ grows with $N$ under both objectives, so the $k=2$ result of Sec.~\ref{sec:results-sweep} is specific to $N=16$. For DFT fitting, the required coupling span is about $30\%$ of the aperture and changes little when the standoff is halved or doubled. For the SIM, the depth saving holds at $N=32$. It shrinks at $N=64$, where the argmax readout shows no saving. Counting tunable elements, connectivity is the cheaper route at $N=16$ ($45$ for the goal-oriented BD-RIS against $64$ for the SIM). Depth is cheaper for goal-oriented designs at $N=32$ ($192$ against $275$) and for both design methods at $N=64$. The reason is that the number of couplings grows as $kN$ and $k^\star$ grows roughly in proportion to $N$. This count ignores the different physical costs of a coupling and a layer.

\subsection{Quantized control}\label{sec:results-quant}
Practical surfaces use few control bits per element (one bit $=$ PIN on/off)~\cite{CuiCoding2014}. We repeat the goal-oriented design of the fully connected BD-RIS and the $L=6$ SIM with $b$-bit control, $b\in\{1,2,3\}$, via quantization-aware training with a straight-through estimator \cite{STE2013}. The forward pass quantizes the configuration ($\B$ for RIS/BD-RIS, phases for SIM), the backward pass updates full-precision shadow parameters, and the best-validated quantized iterate is kept. Each curve uses the best of eight training seeds, selected on a separate validation set.

The right column of Fig.~\ref{fig:results} sweeps SNR per bit depth against the continuous reference. From $20$~dB upward, three-bit control comes within $5\%$ (BD-RIS) and $13\%$ (SIM) of the continuous performance, and the gap grows as the SNR falls. Two-bit control saturates higher, and one-bit control is markedly worse. At one and two bits the result depends on the training run: at two bits and $20$~dB, the BD-RIS RMSE spans $4.4$--$12.8^\circ$ over the eight seeds. These experiments therefore do not separate an expressivity limit from a training limit. At $20$~dB, goal-oriented training improves on DFT-fitted quantization-aware training at one and two bits for both fabrics. At two bits, the selected goal-oriented designs lower the RMSE from $16.4^\circ$ to $4.4^\circ$ (BD-RIS) and from $34.9^\circ$ to $19.8^\circ$ (SIM). At three bits, the selected SIM design gives $2.02^\circ$ against $6.13^\circ$, while the BD-RIS designs and the DFT-fitted baseline are level within the seed spread. Amplitude/phase resolution is only one discretization axis, and denser (sub-$\lambda/2$) sampling typical of PIN metasurfaces is left for future study.

\subsection{Effect of loss}\label{sec:results-loss}
Practical tuners are lossy, in particular PIN diodes. We model a lossy BD-RIS tuner as a shunt conductance $gY_0$ at each port, $\mathbf{Y}=gY_0\I+j\B$, which gives
\begin{equation}
\Th=\big((1-g)Y_0\I-j\B\big)\big((1+g)Y_0\I+j\B\big)^{-1}.
\label{eq:cayleylossy}
\end{equation}
This $\Th$ stays symmetric, but it is a contraction with a state-dependent loss: its magnitude is $(1-g)/(1+g)$ at $\B=\mathbf{0}$ and approaches one as the susceptances grow. For the SIM, a uniform insertion loss IL per layer scales $\W$ by $10^{-L\,\mathrm{IL}/20}$ and thus lowers the SNR at the detectors by $L\,\mathrm{IL}$\,dB. As in \eqref{eq:readout}, the noise enters after the lossy fabric, so the BD-RIS tuner loss also lowers the SNR at the detectors. With loss, $\W$ is not unitary, and the coherent bound also depends on $\W$. We compute it numerically from the output mean $\W\avec(u)$, with the complex gain as a nuisance parameter. Table~\ref{tab:loss} deploys the lossless designs of Sec.~\ref{sec:results-sweep} on lossy hardware. Its lossless columns ($g=0$, $\mathrm{IL}=0$) repeat Table~\ref{tab:budget} and Fig.~\ref{fig:results} within $0.13^\circ$, the Monte Carlo spread between independent noise draws.
\begin{table}[t]
\caption{Lossless designs on lossy hardware ($N=16$, $20$\,dB, noise at the detectors): RMSE in degrees (arg./par.). BD-RIS: tuner loss $g$ and mean power loss of the three designs. SIM: insertion loss IL per layer.}
\label{tab:loss}
\centering
\small
\setlength{\tabcolsep}{9.55pt}
\begin{tabular}{@{}lcccc@{}}
\toprule
BD-RIS: $g$ & $0$ & $0.1$ & $0.2$ & $0.5$\\
loss (dB) & $0$ & $0.7$--$0.8$ & $1.4$--$1.6$ & $2.9$--$3.3$\\
\midrule
$k{=}2$, DFT fit & 7.2/7.0 & 9.1/8.9 & 11.0/10.8 & 14.7/14.4\\
$k{=}2$, goal-or. & 2.7/1.9 & 2.9/2.3 & 3.6/3.1 & 6.4/6.3\\
full, DFT fit & 2.5/1.8 & 2.5/1.8 & 2.6/1.8 & 2.6/2.0\\
\midrule
SIM: IL (dB) & $0$ & $0.5$ & $1$ & $2$\\
\midrule
$L{=}6$, DFT fit & 2.6/1.9 & 2.9/2.2 & 3.4/2.9 & 5.8/5.5\\
$L{=}4$, goal-or. & 2.6/1.8 & 2.6/1.8 & 2.6/1.8 & 2.9/2.2\\
\bottomrule
\end{tabular}
{}
\end{table}
Table~\ref{tab:loss} shows that the fully connected BD-RIS stays near its floor up to $g=0.5$. The sparse $k=2$ surface is more sensitive, because the state-dependent loss distorts its output pattern. For its DFT fit at $g=0.5$, the same power loss applied as a uniform scale gives an argmax RMSE of $8.2^\circ$, against $14.7^\circ$ with the lossy tuners. The goal-oriented design keeps most of its advantage over the DFT fit of the same hardware. At $g=0.2$ its argmax RMSE increases by $32\%$, against $52\%$ for the DFT fit. It does lose parity with the fully connected surface ($6.4^\circ$ against $2.6^\circ$ at $g=0.5$). Retraining through \eqref{eq:cayleylossy}, with the lossless design kept as a candidate, changed the goal-oriented results by at most $0.07^\circ$. At $k=2$, a loss-aware operator fit helps at small loss (argmax: $6.09^\circ$ against $8.15^\circ$ for the lossless fit at $g=0.05$). From $g=0.2$ it hurts, because it trades signal power for operator shape ($36.3^\circ$ against $14.7^\circ$ at $g=0.5$). For the SIM, the DFT-fitted $L=6$ stack leaves its floor at $0.5$\,dB per layer. The goal-oriented $L=4$ and $L=6$ stacks stay on it up to $1$\,dB per layer. The coherent bound stays more than ten times below the RMSE values in Table~\ref{tab:loss}.
{}
\section{Conclusion and Outlook}\label{sec:conclusion}
Treating a programmable surface, its propagation, and its power-only detectors as one physical estimator yields two findings. First, the floors of the standard fit-then-read pipeline are set by the readout grid rather than by the information in the wave. They sit well above the intensity-readout CRB we derive. Second, training the surface through the deployed readout brings sparse hardware to that floor. At $N=16$, a $k=2$ BD-RIS and a four-layer SIM match their fully connected and six-layer operator-fitted counterparts. The DFT-fitted and goal-oriented $k=2$ surfaces carry nearly the same Fisher information, so the gain comes from aligning the response with the readout rather than from acquiring information. The remaining gap to the CRB is a readout limit. The hardware trade-off depends on the array size. Counting tunable elements, connectivity is the cheaper route at $N=16$. Depth becomes cheaper for goal-oriented designs at $N=32$ and for both design methods at $N=64$. The depth saving of goal-oriented training holds up to $N=32$ and shrinks at $N=64$. The connectivity a BD-RIS needs grows with $N$ under both objectives. Equal-budget and cold-start runs show that the objective, rather than the training budget, causes the gain. Under losses of a few decibels, the goal-oriented designs keep most of their advantage over DFT fits of the same hardware. We also mark where training does not reach the floor (e.g., diagonal RIS, two-layer SIMs) and a train/deploy consistency requirement. These findings are established for a linear array in the single-source, single-snapshot, model-matched regime, which we chose on purpose. This choice isolates the effect of the fabric, its complexity, the training objective, and control quantization. The formulation carries over to planar apertures and multi-snapshot readouts. The required hardware must be evaluated again for planar apertures, because it depends on the aperture size (Table~\ref{tab:nscale}). Future work includes these extensions, multi-source scenes with a multi-peak task loss, model mismatch, propagation with aperture spillover, denser apertures, inter-layer reflections, and a general task-sufficient-statistic formulation.

\section*{Acknowledgments}
During the preparation of this work the author used Opus 5.5 (Anthropic) to improve readability and flow of the manuscript, as well as assisting with debugging the simulation codes. The author has rigorously reviewed and edited the final content as needed and takes full responsibility for the claims and results of the manuscript.
{}

\bibliographystyle{IEEEtran}
\bibliography{refs}

\end{document}